\documentclass[twocolumn,letterpaper,superscriptaddress,prl]{revtex4-2}
\usepackage[usenames]{color}
\usepackage{graphicx}
\usepackage{multirow}
\usepackage{verbatim}
\usepackage{amsmath}
\usepackage{amsfonts}
\usepackage{amssymb}
\usepackage{mathrsfs}
\usepackage{url}
\usepackage{color}      
\usepackage{hyperref}   
\usepackage{caption}
\usepackage{float}
\usepackage{amsmath}
\usepackage{braket}
\usepackage{todonotes}
\usepackage[export]{adjustbox}
\usepackage{subfig}

\begin{document}

\title{Giant Optomechanical Coupling in the Charge Density Wave State of Tantalum Disulfide}

\author{Anubhab Haldar}
\email{ahaldar@bu.edu }
\affiliation{Department of Electrical and Computer Engineering, Boston University, Boston, Massachusetts 02215, United States}
\author{Cristian L. Cortes}
\email{ccortes@anl.gov}
\affiliation{Center for Nanoscale Materials, Argonne National Laboratory, Argonne, IL 60439, USA}
\author{Stephen K. Gray}
\email{gray@anl.gov}
\affiliation{Center for Nanoscale Materials, Argonne National Laboratory, Argonne, IL 60439, USA}
\author{Sahar Sharifzadeh}
\email{ssharifz@bu.edu}
\affiliation{Division of Materials Science and Engineering, Boston University, Boston, MA 02215, USA}
\affiliation{Department of Electrical and Computer Engineering, Boston University, Boston, MA 02215, USA}
\author{Pierre Darancet}
\email{pdarancet@anl.gov}
\affiliation{Center for Nanoscale Materials, Argonne National Laboratory, Argonne, IL 60439, USA}
\affiliation{Northwestern Argonne Institute of Science and Engineering, Evanston, IL 60208, USA}

\date{\today}

\begin{abstract}

We study the coupling between light and the structural order parameter in the charge density wave (CDW) state of the layered transition-metal dichalcogenide, tantalum disulfide ($1T-\mathrm{TaS_2}$).
Using time-dependent density functional theory calculations of the dielectric properties along the distortions coordinates, we show that $1T-\mathrm{TaS_2}$ displays a very large change in its dielectric function along the amplitude (Higgs) mode. 
This behavior originates from the coupling of the periodic lattice distortion with an in-plane metal-insulator transition, leading to optomechanical coupling coefficients two orders of magnitude larger than the ones of diamond and ErFeO$_3$. 
In addition, we derive an effective model of the light-induced dynamics, which is in quantitative agreement with experimental observations in $1T-\mathrm{TaS_2}$.
We show that light-induced dynamics along the structural order parameter in $1T-\mathrm{TaS_2}$ can be deterministically controlled to engineer large third-order non-linear optical susceptibilities. 
Our findings suggest that CDW materials are promising active materials for non-linear optics. 
\end{abstract}

\maketitle

Understanding picosecond-scale light-induced dynamics is key for preparation of non-thermal and transient phases with desirable properties for non-linear optics applications \cite{forst2011nonlinear,kampfrath2013resonant,lloyd20212021,sie2019ultrafast,baldini2019exciton}. 
Layered transition metal dichalcogenides (TMDC) and their heterostructures, owing to their strong coupling with light, have been at the forefront of this search for novel transient electronic properties \cite{jin2018ultrafast}.

Experimental observations of the ultrafast laser-induced dynamics in the low-temperature, broken-symmetry phases of the layered TMDC $1T \mathrm{TaS_2}$ \cite{demsarFemtosecond2002,perfettiTime2006,eichbergerSnapshots2010,deanPolaronic2011,hellmannTime2012,han2015exploration,avigoExcitation2018,zongUltrafast2018} have shown a strong coupling between light and structural distortions. 
In particular, a defining feature in these experiments is the strongly non-thermal response of the material, characterized by a selective, coherent excitation of the amplitude (Higgs) mode, that displaces the system from its low-symmetry phase to its high-symmetry counterpart. 
The excitation of the Higgs mode has been observed using different techniques, including time-resolved absorption/reflection \cite{demsarFemtosecond2002,deanPolaronic2011}, ultrafast electron diffraction and microscopy \cite{eichbergerSnapshots2010,zongUltrafast2018}, and time-resolved photoemission spectroscopy \cite{perfettiTime2006,hellmannTime2012,avigoExcitation2018} for excitation energies varying from 0.62 eV \cite{deanPolaronic2011} to 3.2 eV \cite{eichbergerSnapshots2010}, suggesting a general, non-resonant coupling, and is also reproduced by first-principles calculations \cite{zhangPhotoexcitation2019}. Thus, $1T-\mathrm{TaS_2}$ is a  prototypical material for study of non-equilibrium phases of matter \cite{han2015exploration,stojchevska2014ultrafast}.

However, to the best of our knowledge, no microscopic mechanism explaining the coupling of structural order parameters and light \cite{sun2020transient} has been proposed in $1T-\mathrm{TaS_2}$, unlike other materials such as multiferroics \cite{subedi2014theory,juraschek2017dynamical}. As $1T-\mathrm{TaS_2}$ is one of many known charge density wave (CDW) materials, this lack of a microscopic description also raises the question of the universality of the light-induced dynamics observed in $1T-\mathrm{TaS_2}$  for this class of materials. 

In this work, we show how the light-induced dynamics in the CDW state of $1T-\mathrm{TaS_2}$ results from the coupling of light with the  \emph{change} in the dielectric function along its structural Higgs mode. 
Using extensive density functional theory (DFT) and time-dependent density functional theory (TDDFT) calculations, we compute the frequency-dependent polarizability of $1T-\mathrm{TaS_2}$ along its structural Higgs and Goldstone coordinates.  We show that the coupling of lattice distortions with a metal-insulator transition leads to very large optomechanical coupling coefficients, at least two orders of magnitude larger than the ones of diamond, BiFeO$_3$, and ErFeO$_3$. Using our TDDFT results, we derive an effective classical model of the light-induced dynamics of the structural Higgs and Goldstone modes in CDW materials, in quantitative agreement with experimental observations of light-induced dynamics in $1T-\mathrm{TaS_2}$~\cite{deanPolaronic2011,hellmannTime2012}.
We show how these effects enable the light-induced dynamics of the Higgs mode to be deterministically controlled in order to engineer large non-linear optical susceptibilities in materials that couple periodic lattice distortions with metal-insulator transitions.

 At temperatures lower than 180 K, $1T-\mathrm{TaS_2}$, along with the isovalent $1T-\mathrm{TaSe_2}$, exhibits a so-called ``Star-of-David'' commensurate CDW involving an in-plane, $\sqrt{13}\times\sqrt{13}R=12.4^{\circ}$ periodic-lattice distortion, associated with an insulating state~\cite{dardelTemperaturedependent1992,perfettiSpectroscopic2003,perfettiTime2006,siposMott2008,kimObservation1994,stoltzTunneling2007,colonnaMott2005} -- unlike the metallic behavior of the high-temperature undistorted phase~\cite{disalvoLow1977,wilsonTransition1969, wilsonChargedensity1975,fazekasElectrical1979,fazekasCharge1980}. We note that the nature of this low-temperature insulating state, possibly a Mott insulator \cite{fazekas1979electrical,fazekasCharge1980,perfettiSpectroscopic2003,perfettiTime2006,perfettiFemtosecond2008,ma2016metallic,law20171t} or an in-plane band insulator with Anderson localization out-of-plane caused by packing disorder~\cite{disalvoLow1977,dardelTemperaturedependent1992,bovet2003interplane,DarancetPhysRevB.90.045134,ritschel2018stacking,lee2019origin,PhysRevLett.126.196405}, is still a matter of contemporary debate, and beyond the scope of this work. While differing on the nature of the out-of-plane character, both scenarios are compatible with the DFT description of the in-plane insulating band-structure~\cite{DarancetPhysRevB.90.045134,ma2016metallic}, which is in excellent agreement with angle-resolved photoemission spectroscopy~\cite{bovet2003interplane,lee2019origin}, vibrational spectroscopy~\cite{albertini2016zone}, and time-resolved spectroscopy~\cite{zhangPhotoexcitation2019} experiments.
 
 In our work, all configurations are modeled using a $\sqrt{13}\times\sqrt{13}\times1$ unit cell with $13$ Tantalum atoms, and a vertical stacking of center of distortions~\cite{bovet2003interplane,suzuki2004electronic,ge2010first,DarancetPhysRevB.90.045134}. In the ground state, this unit cell is conducive to the in-plane insulating, and out-of-plane metallic band-structure described above~\cite{bovet2003interplane,suzuki2004electronic,ge2010first,DarancetPhysRevB.90.045134}, and, for this reason, we restrict our study to the in-plane response.

We compute the total energy and the components of the macroscopic longitudinal dielectric function tensor, $\epsilon (\mathbf{q} \rightarrow  \mathbf{0};  \omega)$, using the GPAW implementation~\cite{GPAW} of DFT~\cite{GPAW_DFT} and TDDFT~\cite{GPAW_TDDFT}, respectively. Full computational details are given in the Supporting Information. We use the generalized gradient approximation of Perdew, Burke, and Ernzerhof (PBE)~\cite{perdew1996generalized} with an effective Hubbard U value of 2.27 eV~\cite{DarancetPhysRevB.90.045134}, with 5 (6) valence electrons for Ta (S). TDDFT calculations are performed within the random phase approximation (RPA). We use an inhomogeneous configuration grid of 188 configurations for interpolations of the total energy and dielectric functions.

\begin{figure}[t]
	\centering
    \includegraphics[width=\linewidth]{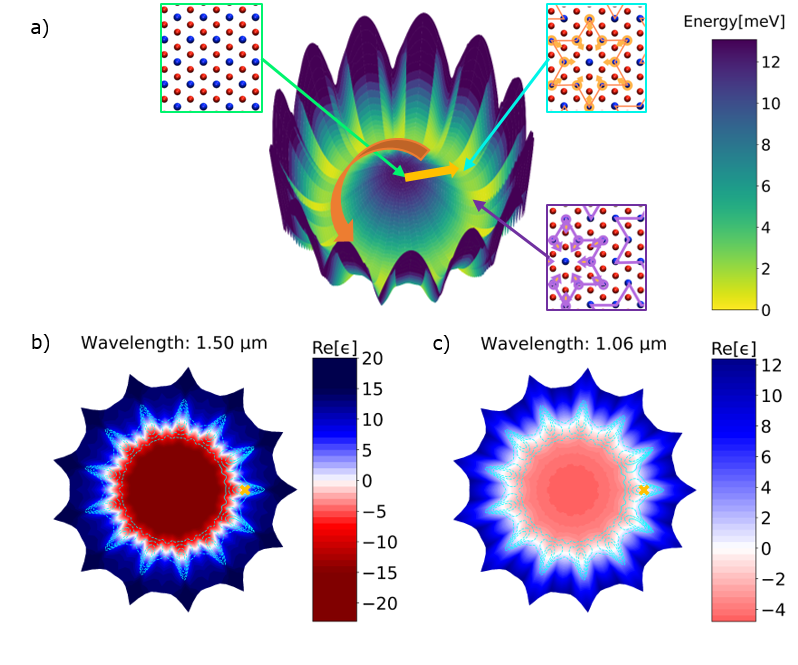}
	\caption{a) Total energy of $1T-\mathrm{TaS_2}$ with respect to the Higgs and Goldstone coordinates. The range of displacement along the radial coordinate is 1.30 Å. The energy is indicated per TaS$_2$.The minima along the Goldstone directions correspond to a Tantalum atom at the center of the displacement field. b) and c) Corresponding real part of the in-plane dielectric function ($\epsilon_{||}$) at two different excitation wavelengths. Cyan isoenergy contours are placed 0.77 meV / TaS$_2$ apart, starting at 0.77 meV.}
    \label{fig:f1}
\end{figure}

We only consider structural distortions along the order parameter of the CDW. As the CDW is associated with a symmetry breaking, the problem is (at least) two-dimensional with an amplitude (Higgs) and a phase mode (Goldstone), as shown in Fig. \ref{fig:f1} (a). In the context of periodic lattice distortions, the Higgs coordinate corresponds to the amplitude of the vector field describing atomic displacements between the high symmetry (1 Ta atom per unit cell) and low symmetry (13 Ta atom per unit cell) configurations. Correspondingly, the Goldstone coordinate refers to the origin of this vector field within the unit cell. As solids lack continuous symmetry, energy fluctuates along the Goldstone coordinate. In $1T-\mathrm{TaS_2}$, the ground-state CDW configuration is associated with a Ta atom at the center of the distortion, hence the energy has 13 distinct minima along the Goldstone coordinate. As is customary, the Higgs (Goldstone) coordinate is represented by the radial $r$ (angular $\theta$) direction. 

Fig. \ref{fig:f1} (a) shows the interpolated DFT total energy along these coordinates. As expected, the extrema of the total energy include the 13 distinct energy minima at $r_{\mathrm{eq}}\approx0.77$Å and $\theta\equiv 0 \mod 2\pi/13 $, with a local maximum associated with the high symmetry cell ($r=0$) 12.7 meV/$\mathrm{TaS_2}$. Additionally, saddle points of energy 2.65 meV/$\mathrm{TaS_2}$ are found at ($r=$ 0.65Å, $\theta\equiv \pi/13 \mod 2\pi/13 $) using the nudged elastic band method, confirming that the Goldstone mode acquires a finite energy due to the discrete lattice symmetry. 

As mentioned above, in $1T-\mathrm{TaS_2}$, the periodic lattice distortion is associated with an in-plane metal-insulator transition. Accordingly, and as shown in Fig. \ref{fig:f1} (b), the real part of the dielectric function $\mathrm{Re}\left[\epsilon(\omega;r,\theta)\right]$ of $1T-\mathrm{TaS_2}$ is negative near the high symmetry cell, $r=0$, and becomes positive at values of $r$ near the ground state ($r_{\mathrm{eq}}\approx0.77$Å). While this general trend is observed for all  frequencies considered in this work ( 0.5 eV $< \omega <$ 2.0 eV, as shown in the SI),  the $r$ value at which the transition occurs $\mathrm{Re}\left[\epsilon(\omega;r,0)) \right]= 0$ increases as  $\omega$ increases (Fig. \ref{fig:f1} (c)). Moreover, the amplitude of the change along the Higgs coordinate, $| \mathrm{Re}\left[\epsilon(\omega;0,0)-\epsilon(\omega;r_{\mathrm{eq}},0)\right] |$, decreases as $\omega$ increases, a consequence of higher energy transitions having a weaker dependence on the atomic details of the lattice due to their higher kinetic energy.  

Importantly, for all $\omega$, the relative variations of  $\mathrm{Re}\left[\epsilon(\omega;r,\theta)\right]$ along $r$ are very large, with values  $| \mathrm{Re}\left[\epsilon(\omega;0,0)-\epsilon(\omega;r_{\mathrm{eq}},0)\right]/\mathrm{Re}\left[ \epsilon(\omega;r_{\mathrm{eq}},0)\right]  | >>  1 $. In contrast, the variations of  $\mathrm{Re}\left[\epsilon(\omega;r,\theta)\right]$ along  $\theta$, while finite, are found to be on the order of $\mathrm{Re}\left[\epsilon(\omega;r;\theta=\pi/13)\right] - \mathrm{Re}\left[\epsilon(\omega;r;\theta=0)\right] < 5$.

\begin{figure*}[t]
    \centering
    \includegraphics[width=0.9\textwidth]{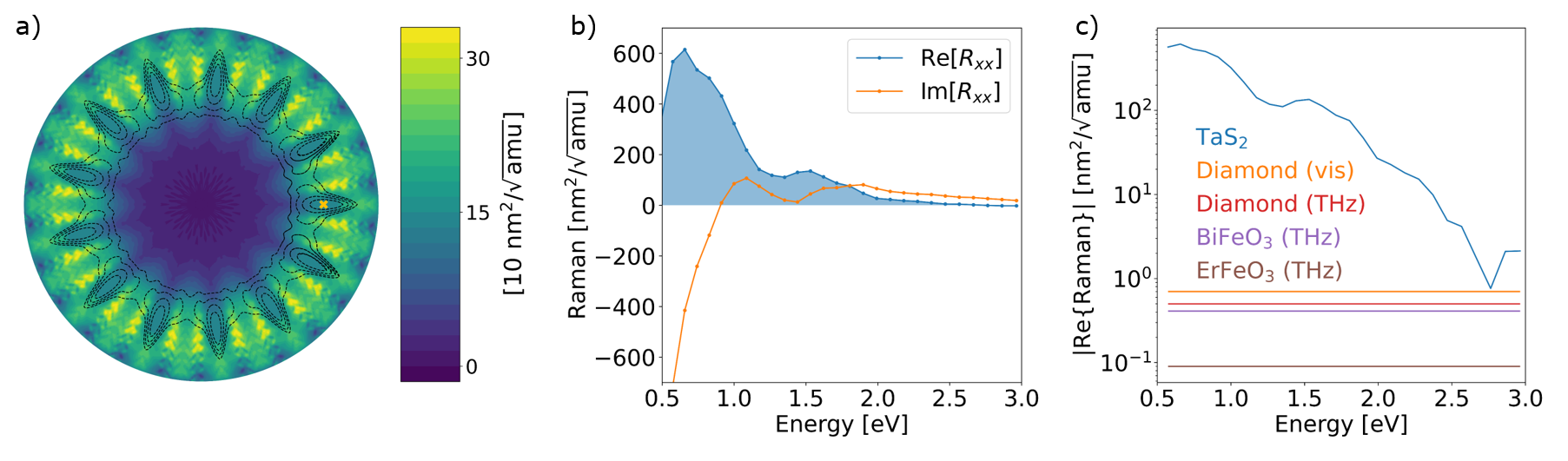}
    \caption{a) The Raman cross-section with the amplitude mode for incoming light at $\omega =$ 1.17 eV $\frac{\partial \alpha(\omega;r,\theta)}{\partial r} \left(r,\theta\right)/\sqrt{m}$. Black isoenergy contours are placed 0.77 meV / TaS$_2$ apart, starting at 0.77 meV. b) Real and imaginary parts of the Raman tensor at $(r_{\mathrm{eq}},\theta=0)$ of the amplitude mode as a function of light frequency. c) Comparison of $1T-\mathrm{TaS_2}$ Raman (this work) with other materials. The values for diamond, BiFeO$_3$, and ErFeO$_3$ are taken from ref. \cite{juraschekSumfrequency2018}.} 
    \label{fig:f2}
\end{figure*}

This variation of $\epsilon(\omega;r,\theta)$ along the Higgs coordinate has a large impact on the Raman activity, i.e. the Raman tensor, and is related to the optomechanical coupling coefficient \cite{gonze1997dynamical,juraschek2018sum}. We can extract this coefficient from direct differentiation of the RPA polarizability $\alpha(\omega;r,\theta)$, computed using TDDFT, along the distortion coordinates, as shown in Fig. \ref{fig:f2} (a). The optomechanical coupling between light and the Higgs order parameter, $m^{-1/2}\frac{\partial \alpha(\omega;r,\theta)}{\partial r} \left(r,\theta\right)$ (where $m$ is the effective mass of the mode), is maximal near the CDW ground-state with $r = r_{\mathrm{eq}}$, progressively vanishing near the high symmetry phase as $r \rightarrow 0$. This suggests that, in $1T-\mathrm{TaS_2}$, light can only transiently stabilize the high-symmetry phase, and that the amplitude mode cannot be excited from the high-symmetry phase. Moreover, the optomechanical coupling near $r_{\mathrm{eq}}$ strongly depends on $\theta$, suggesting that a protocol of excitation involving the Goldstone mode~\cite{juraschek2020parametric} could further enhance the optomechanical coupling. 

In agreement with the larger dielectric changes at lower frequencies (Fig. \ref{fig:f1}), the optomechanical coupling at $\left(r=r_{\mathrm{eq}},\theta=0 \right)$ decreases at higher frequencies (Fig. \ref{fig:f2} (b)). Moreover, we note a large increase in the magnitude of imaginary part of the Raman tensor at $\omega \lesssim 1.eV$. This imaginary part is related to the two-photon absorption/emission of the materials, suggesting $1T-\mathrm{TaS_2}$ could be used as a platform for entangled photon emission. 

The large change in dielectric function ($\Delta \mathrm{Re}\left[\epsilon\right] \approx 50$) over a small displacement ($\Delta r < 1$\AA ) enabled by the charge density wave results in a giant value of the optomechanical coupling coefficient, shown in Fig. \ref{fig:f2} (c). At optical frequencies, the computed values are two orders of magnitude larger than $\mathrm{BiFeO}_3$, $\mathrm{ErFeO}_3$ and diamond \cite{juraschekSumfrequency2018}. We note that the metallic nature of the CDW phase predicted in DFT prevents us from modeling accurately the polarizability of the system at $\omega\lesssim 1/\tau$, the Drude damping limit. Nevertheless, a fully insulating CDW state would lead to even \emph{larger} optomechanical coupling strength at lower frequencies. 

We conclude this work by discussing the significance of our findings and of the large optomechanical coupling in a CDW materials in the context of the search for  materials with large third-order non-linear susceptibility  $\chi^{(3)} (\omega_1, \omega_2, \omega_3)$. In particular, $\chi^{(3)} (\omega_1, \omega_2, \omega_3) \propto |R(\omega_1)|^2 H (\omega_2, \omega_3)$~\cite{maehrlein2017terahertz}, where $|R|^2$ is the square of the Raman tensor plotted in Fig.~\ref{fig:f2}, and $H(\omega, \omega')$ is the phonon propagator. The large values of $|R|$ found above (Fig.~\ref{fig:f2} (c)) suggests that $1T-\mathrm{TaS_2}$ and other CDW materials could be promising for non-linear optics applications. Nevertheless, a prerequisite to using CDW materials for non-linear optics is the generation of deterministic trajectories of the system with light, across configurations with large dielectric constant contrast.

To analyze this possibility, we use a classical equation of motion describing the Higgs and Goldstone coordinates coupled with light through Raman scattering derived in~\cite{subedi2014theory,juraschek2017ultrafast,juraschek2017dynamical,juraschekSumfrequency2018}:
\begin{equation}
    m \Ddot{\mathbf{X}} = - \frac{\partial U}{\partial \mathbf{X}} - \gamma \dot{\mathbf{X}}  + R\mathbf{E}^2,
    \label{Eq:THzDynamics}
\end{equation}
where $\mathbf{X}=\left(r,\theta\right)$, $U$ is the total energy computed in DFT, $\gamma$ is an empirical damping parameter, and $\mathbf{E}^2$ is the electric field associated with an in-plane polarization. Details of the implementation of this equation on our inhomogeneous grid are given in the supporting information. While the forces (Raman tensor) are computed and interpolated from DFT (TDDFT) at each $\left(r,\theta\right)$, we note that this classical description based on the Born-Oppenheimer approximation is not suitable for the description of the sub-picosecond dynamics~\cite{shen2014nonequilibrium,wang2018theoretical,freericks2009theoretical} and describes the system evolution at times larger than the timescales associated with electronic thermalization~\cite{sadasivam2017theory,kemper2018general}.

\begin{figure}[t]
	\centering
	\includegraphics[width=\linewidth]{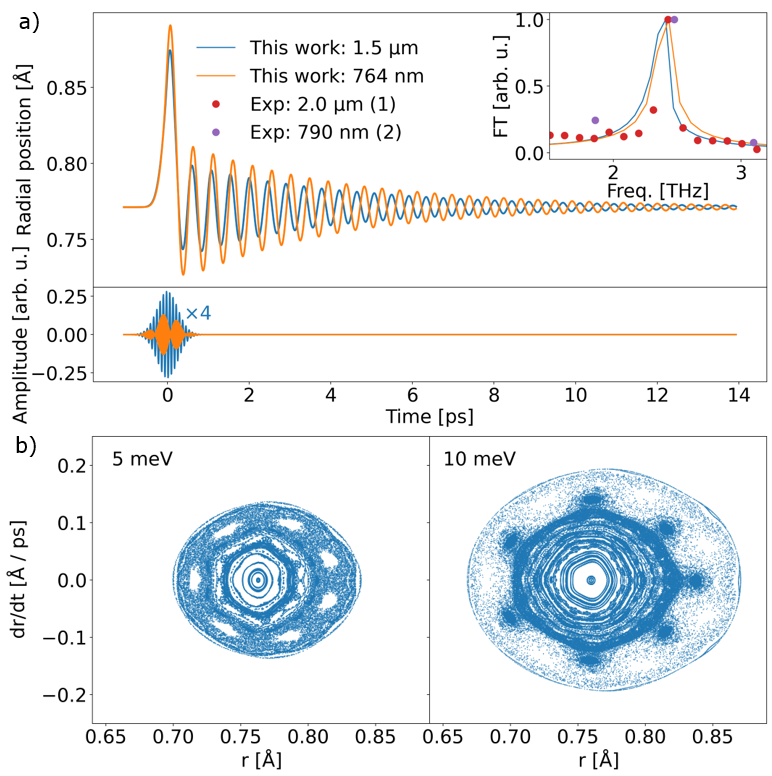}
    \caption{a) Time-evolution of the amplitude mode after excitation by short optical pulses of wavelengths (time profile of the pulse shown below) predicted by Eq.~\ref{Eq:THzDynamics} (empirical damping parameter $\gamma=0.5 $THz). Inset: Fourier transform of the damped response after the initial transient response has subsided and experiments, where (1) is taken from Ref. \cite{deanPolaronic2011} and (2) from Ref. \cite{hellmannTime2012}. b) Poincaré surfaces of sections at different energies (Left: 5meV/13Ta; Right 10meV/13Ta) of the conjugate variables $r$ and $\dot{r}$ (corresponding to movement and position along the Higgs mode) when crossing the $\dot{\theta}=0$ plane for the pure potential (no damping nor external field). }
    \label{fig:f3}
\end{figure}

In Fig. \ref{fig:f3} (a), we show the time-evolution subsequent to the application of a light pulse, computed using Eq. \ref{Eq:THzDynamics}. As the Raman tensor in Eq. \ref{Eq:THzDynamics} depends on the frequency of the pulse, $\omega$, the coupling between light intensity and dynamics is also explicitly dependent on $\omega$, with lower values of $\omega$ resulting in stronger coupling at constant intensity. 

The predicted time-evolution computed using a pulse intensity of $5\times10^{-3}$ V/Å and $2\times10^{-2}$ V/Å for the infrared and visible pulse respectively, is in quantitative agreement with experimental values~\cite{perfettiTime2006,eichbergerSnapshots2010,deanPolaronic2011,hellmannTime2012,avigoExcitation2018,demsarFemtosecond2002,sugaiLattice1985}, further validating the proposed non-resonant coupling mechanism between CDW order parameter and light through Raman processes as the mechanism underlying light-induced dynamics in $1T-\mathrm{TaS_2}$. Moreover, we note that our proposed coupling is also consistent with recent observations of an unusual intensity-dependence of the optical properties of $1T-\mathrm{TaS_2}$ in its incommensurate phase~\cite{li2020large}. Importantly, the low energy of the saddle points ($\approx 35$ meV per unit cell of 13 Ta atoms) along the Goldstone coordinate suggests a possible chaotic dynamics even at low pulse energy. 

To test the stability of the dynamics necessary to generate large non-linear susceptibilities through optomechanical coupling, we construct the Poincaré surface of sections (details in Supplementary Information) using orbits generated from Eq. \ref{Eq:THzDynamics} without dissipation or coupling with light. As shown in Fig. \ref{fig:f3} (b), at excess energies as low as 5 meV and 10 meV per unit cell, the phase space displays some regions associated with chaotic trajectories and regular trajectories separated by Kolmogorov–Arnold–Moser tori. The chaotic regions are associated with large fluctuations in $r$ (i.e. small deviations in the $\theta$ directions). Regular orbits exist for lower fluctuations in $r$ around $r_{\mathrm{eq}}$,  on the scale of $r_{\mathrm{eq}}\pm 0.05$Å Interestingly, this suggests that excitation of the system along the Goldstone coordinate is needed to generate regular orbits with Higgs fluctuations. At higher energies, the phase space is still mixed, with the areas of chaotic and regular orbits both expanding. This indicates that an initial excitation protocol of the system along both Higgs and Goldstone coordinates may be necessary for harnessing the giant optomechanical effect in  $\chi^{(3)}$  applications.

In conclusion, we demonstrated that the coupling between light and the structural order parameter in the CDW state of $1T-\mathrm{TaS_2}$ is mediated by the change in dielectric function along the amplitude (Higgs mode). Using extensive DFT/TDDFT calculations, which we showed to be in quantitative agreement with time-resolved experiments, we revealed  that this optomechanical coupling is larger than in any other known material, due to the coupling between the periodic lattice distortion and a metal-insulator transition. Our results strongly suggest CDW materials are promising active media for non-linear optics applications, provided the system is carefully prepared in a mixture of a Higgs and Goldstone modes.

 This work was performed, in part, at the Center for Nanoscale Materials, a U.S. Department of Energy Office of Science User Facility, and supported by the U.S. Department of Energy, Office of Science, under Contract No. DE-AC02-06CH11357. A.H. and S.S. acknowledge financial support from the U.S. Department of Energy (DOE), Office of Science, Basic Energy Sciences Early Career Program under Award No. DE-SC0018080. This material is based upon work supported by Laboratory Directed Research and Development (LDRD) funding from Argonne National Laboratory, provided by the Director, Office of Science, of the U.S. Department of Energy under Contract No. DE-AC02-06CH1135. P.D. was supported by the U.S. Department of Energy, Office of Science, Office of Basic Energy Sciences Data, Artificial Intelligence and Machine Learning at DOE Scientific User Facilities program under Award No. 34532.

%

\end{document}